\documentclass[journal]{vgtc}              
\DeclareUnicodeCharacter{202F}{~}    
\DeclareUnicodeCharacter{0096}{~}   
\usepackage[normalem]{ulem}
\usepackage{graphicx,multirow}
\usepackage[font=small,skip=5pt]{caption}

\onlineid{0}

\vgtccategory{Research}

\title{Exploring the Effect of Viewing Attributes of Mobile AR Interfaces on Remote Collaborative and Competitive Tasks}

\author{%
  \authororcid{Nelusha Nugegoda}{0009-0007-8095-542X},
  \authororcid{Marium-E- Jannat}{0000-0002-4815-4741}, 
  \authororcid{Khalad Hasan}{0000-0002-4815-5461}, and \authororcid{Patricia Lasserre}{0000-0001-7080-2437}
}

\authorfooter{
  \item
  	Nelusha Nugegoda (Corresponding Author) is with the Department of Computer Science, University of British Columbia, Canada. E-mail:nelusha.nugegodawattaranthenne@ubc.ca
  \item
  	Marium-E- Jannat is with the Department of Computer Science, University of British Columbia, Canada. E-mail:marium.jannat@ubc.ca
  \item 
  Khalad Hasan is with the Department of Computer Science, University of British Columbia, Canada. E-mail: khalad.hasan@ubc.ca
  \item 
  Patricia Lasserre is with the Department of Computer Science, University of British Columbia, Canada. E-mail:patricia.lasserre@ubc.ca
}

\newcommand{\update}[1]{\textcolor{black}{#1}}

\abstract{
Mobile devices have the potential to facilitate remote tasks through Augmented Reality (AR) solutions by integrating digital information into the real world. Although prior studies have explored Mobile Augmented Reality (MAR) for co-located collaboration, none have investigated the impact of various viewing attributes that can influence remote task performance, such as target object viewing angles, synchronization styles, or having a secondary small screen showing other users current view in the MAR environment. In this paper, we explore five techniques considering these attributes, specifically designed for two modes of remote tasks: collaborative and competitive. We conducted a user study employing various combinations of those attributes for both tasks. In both instances, results indicate users' optimal performance and preference for the technique that allows asynchronous viewing of object manipulations on the small screen. Overall, this paper contributes novel techniques for remote tasks in MAR, addressing aspects such as viewing angle and synchronization in object manipulation alongside secondary small-screen interfaces. Additionally, it presents the results of a user study evaluating the effectiveness, usability, and user preference of these techniques in remote settings and offers a set of recommendations for designing and implementing MAR solutions to enhance remote activities.

} 


\graphicspath{{figs/}{figures/}{pictures/}{images/}{./}} 
\usepackage{tabu}                      
\usepackage{booktabs}                  
\usepackage{lipsum}                    
\usepackage{mwe}                       
\usepackage{amssymb}
\usepackage{mathptmx}                  

\begin{document}



\maketitle


\section{Introduction}

Mobile Augmented Reality (MAR) solutions are effortlessly integrating virtual content into the real world \cite{Perdue2020:TrainingData} via smart devices like smartphones. With over 60\% of the global population using mobile phones \cite{bankmycell_2024}, MAR solutions are rapidly gaining popularity and becoming widespread. This widespread access to augmented digital information in real-world settings has broadened the scope of domains such as interactive learning \cite{10.1145/2669711.2669870, su16031192, Lah2024:MAR_Chemistry}, design \cite{10.1023/A:1023084706295, Wang2019MultimodalInteraction, WANG2010778}, and fundamentally reshaped our interaction with digital information. For instance, MAR has enabled the detailed study of human anatomy by overlaying complex anatomical structures onto real-world environments \cite{10.1145/2669711.2669870}, facilitating an immersive learning experience. Similarly, MAR applications support learning in facial reconstructive surgeries \cite{GUO2024107970} and oral health education \cite{ROMALEE2024105277}, offering immersive experiences that contribute to the practical training of medical professionals and health education for community-dwelling older adults. Additionally, the application of MAR in learning chemistry subjects \cite{Lah2024:MAR_Chemistry} demonstrates its capability to enrich students' comprehension of scientific concepts through interactive and engaging virtual experiments.

In recent years, researchers showed that Augmented Reality (AR) interfaces have the potential to facilitate collaborative activities \cite{billinghurst1998shared}. For instance, Henrysson et al. \cite{henrysson2005face} were one of the earlier researchers exploring the usability of co-located collaboration using a face-to-face mobile game. Wells et al. \cite{10.1145/3313831.3376541} explored how MAR can support co-located collaborative activities. They conducted an observational study to underscore the need for design considerations that reduce cognitive and physical demands and enhance collaborative dynamics. Similarly, Nilsson et al. 
\cite{nilsson2009using} discussed how AR can be used in fostering collaboration among emergency services, revealing positive user feedback and performance gains over joint planning tasks using a simulated emergency scenario. They also highlighted the significance of iterative design in creating effective collaborative tools for crisis management scenarios.

Wells et al. \cite {Wells2022:Co-Located_Collaboration} further investigated how different mobile device configurations affect co-located collaboration. They observed different collaborative strategies employed by participants based on different device sizes and configurations. 
Additionally, Project IRL explored augmenting in-person social interactions through five mobile apps, addressing the critique of digital technology's isolating effects by fostering playful, co-located interactions. Through a user study with 101 participants, they offered design guidelines focused on device arrangement, enablers, modifying reality's affordances, and enhancing co-located play, aiming to guide future social AR experiences\cite{10.1145/3512909}.  
Even though some efforts have been made to explore co-located collaborative activities through MAR, there remains a significant gap in understanding the dynamics of remote tasks, both collaborative and competitive, when all the participants are not physically co-present.


Mobile devices allow users to collaborate from anywhere at any time, providing flexibility to work together without the constraints of co-located physical proximity. In recent years, the increasing adoption of remote work and distributed teams has made remote tasks through mobile phones more prevalent. However, completing such tasks on small-screen devices presents unique challenges, such as real-time coordination and spatial awareness of activities. To enhance the remote activity experience, we investigated various factors such as content viewing angles for improved spatial awareness, enabling both synchronous and asynchronous manipulation of virtual objects to accommodate different collaboration styles, and employing a secondary screen to display each other's activities for enhanced spatial awareness. We designed five distinct techniques using these attributes and evaluated their effectiveness in enhancing the remote task experience for both collaborative and competitive tasks. To the best of our knowledge, prior studies have not explored the influence of these factors on both collaborative and competitive tasks in remote MAR settings.

We conducted a user study to investigate how these attributes influence users' behavior, strategies, and performance in remote MAR settings. The study was conducted with 10 groups of 2 participants where they were required to perform a set of collaborative and comparative tasks remotely. Results from our study showed that the most preferred technique for both collaborative and competitive tasks is one that doesn't involve synchronized manipulations but allows participants to view each other's screens through a secondary small screen on the interface. Based on the results, we developed a set of design recommendations for future MAR solutions to provide an enhanced experience of remote collaboration and competition.

This paper makes the following contributions: (i) Design of five techniques for remote collaborative and competitive tasks in MAR including attributes such as viewing angle, synchronization in object manipulation, and having a secondary small screen interfaces; (ii) A comprehensive user study that assesses the effectiveness, usability, and user preference regarding the five techniques in remote settings for both collaborative and competitive tasks; and (iii) A set of recommendations for the design and implementation of MAR solutions that facilitate effective and engaging remote collaboration and competition.

\section{Related Work}
Our study examines how different viewing attributes influence the performance of remote collaborative and competitive MAR tasks. This section reviews the prior works that investigated different dimensions of multi-user engagement and collaboration through AR applications.

\subsection{Collaboration in Augmented Reality} 

Collaborative activities can be categorized into co-located and remote activities where all participants join together to accomplish a shared goal \cite{10.5555/524573,8456568,kim2018effect,kim2020combination,kim2020multimodal}. In recent years, researchers have also leveraged AR technology to support collaborative tasks. For instance, Bruno et al. \cite{bruno2019augmented} present an AR tool that allows efficient data collection and information exchange between different workers involved in the design and production process of the oil and gas industry. 
Van Lopik et al. \cite{VANLOPIK2020103208} designed an Augmented Repair Training Application for shop floor users for AR content creation. They identified the need for user-friendly templates to overcome the complexity of AR content creation and promote industrial adoption of AR within \update{Small Enterprises (SE).}

Recently, there has been a notable increase in the adoption of technology and design for enhancing co-located collaborative efforts \cite{lundgren2015designing,fischer2018beyond}. Lundgren et al. \cite{lundgren2015designing} introduced a design framework to guide the development of applications to facilitate co-located activities with mobile devices. 
Knoll et al. \cite{10.1145/3544549.3585841} introduced an escape room designed for collaborative play by two players. They conducted semi-structured interviews with four dyads to explore participants' interpersonal dynamics and experiences during gameplay. Results revealed that participants found the experience collaborative and enjoyable, as it promoted discussion and fostered social dynamics. However, participants sometimes felt disoriented by the virtual content.
Koceski et al. \cite{10.1007/978-3-642-28664-3_13} investigated how AR board games affect collaborative thinking and social interaction. They investigated the social interaction between players and their performance and the usability of mobile phones for collaborative AR. Results demonstrated that AR applications can naturally enhance players' interaction and increase their confidence compared to other gaming conditions. Other studies \cite{10.1145/3313831.3376541, Wells2022:Co-Located_Collaboration} have highlighted the potential of AR to enhance co-located collaboration through shared digital information overlays and interactive 3D models on mobile devices.

Remote collaboration can introduce complexities in a seamless collaborative experience as physical absence poses challenges in providing a smooth and intuitive collaborative experience. Prior work explored AR-based solutions to enable seamless and collaborative experiences with virtual objects among remote users  \cite{kim2018revisiting,hall2018capturing,wang2016comprehensive}.
Marques et al. \cite{10.1016/j.cag.2021.08.006} emphasized the importance of enhanced characterization and evaluation in remote collaboration-based AR solutions. They introduced a roadmap to outline essential factors for evaluating such solutions. They suggested that remote collaboration mediated by AR is currently positioned between the Replication and Empiricism phases of the BRETAM model \cite{10.5555/106027.106028}. To address this, they proposed a roadmap outlining critical research actions necessary to facilitate a more comprehensive understanding of the collaboration process in future AR solutions.
Prior work has primarily focused on collaborative activities in AR, leaving a critical gap in exploring how MAR can facilitate a seamless and intuitive remote experience, both collaboratively and competitively, on a virtual object shared by remote users.

\subsection{Cross-Device Remote Interaction}
Though there is considerable work on remote collaboration using AR, the majority of the solutions are built upon multiple device interfaces. Billinghrust and Kato \cite{10.1145/632716.632838} presented a remote video conferencing solution using AR markers where the local user scans a marker with their HMD, which then displays a video feed from the remote user's desktop augmented into the local user's HMD. Johnson et al. \cite{10.1145/2675133.2675176} compared HMD and tablet to explore the impact on both devices for collaboration. They used an interactive phone call as a synchronous remote collaborative system for collaborative video calls. They observed that in dynamic tasks, the users with HMDs offer more frequent directing commands with proactive assistance. However, in static tasks, researchers observed conflicting perceptions on how to use the technologies successfully. 

Seo et al. \cite{10.1145/3567561} designed HoloMeter which is an HMD-based remote collaborative system where a local worker wears AR HMD while performing a physical task and a remote helper interacts with the local user drawing through an Actionpad (i.e., tablet or mobile phone) while observing from their computer. By conducting a user study, they showed that this innovative framework supports effective and efficient communication between remote users. Further, they suggest future research could explore its scalability in complex, multi-user scenarios to understand its impact on remote collaborative tasks. Fink et al. \cite{10.1145/3567709} developed Re-locations, a system that facilitates remote collaboration in incongruent spaces. The authors demonstrated that this system reduces the complexity of incongruent physical and spatial workspaces by enabling users to interact collaboratively in relation to user-defined virtually shared landmarks via HMDs, while a local user interacts with their desktop computer. Although there is a significant amount of research focused on remote systems across various cross-device systems, it is critical to study how using multiple mobile phones for remote activities affects users' overall experience.
  
\subsection{Design Considerations in Digital Collaboration}

Researchers highlighted the key design challenges to the success of collaborative environments such as high awareness of others' actions and intentions, high control over the interface, high availability of background information, and information transparency among collaborators while preventing user privacy leakages \cite{10.1145/2147783.2147784,10.1145/1873561.1873568}. Studies have focused on using virtual annotations using drawings, hand gestures, 2D images or live streaming, pointers, and eye gaze \cite{lukosch2015collaboration,kim2018revisiting,kim2020multimodal,ens2019revisiting,de2019systematic}. In addition to these standard virtual annotations, recent studies focused on using virtual replicas \cite{10.1145/2807442.2807497, 10.1145/3089269.3089281,bai2012analytic} and 3D reconstruction of physical environments \cite{10.1016/j.intcom.2012.07.004, ANTUNES2014146} to give a better understanding of other users' interactions, mitigating the confusion that occurs when manipulating virtual objects.  

Unlike co-located settings where participants can rely on natural cues and direct interaction with shared objects, remote collaboration requires innovative solutions to replicate these dynamics at a distance \cite{10.1016/j.cag.2021.08.006}. The key challenges include maintaining a sense of presence, ensuring real-time communication, and providing mutual awareness of actions and changes within the shared digital environment. 
\update{Stoakley et al. \cite{10.1145/223904.223938} introduced the ``Worlds in Miniature (WiM)'' metaphor where users can interact with a virtual environment using a hand-held miniature version of the virtual environment in a head-tracked display. The proposed system provides the users the ability to see a miniature representation of the virtual world as well as the ability to perform selection, manipulation, travel, and visualization tasks.}
Gauglitz et al. \cite{10.1145/2642918.2647372} designed a system for remote collaboration where they introduced a small screen to display the remote user's view, highlighting the importance of having a dedicated screen to view others' view. Here, the remote user used a tablet while the local user worked on a PC. Fages et al. \cite{10.1145/3555607} explored several views that remote desktop users could utilize, including a fully virtual representation, a first-person view, and an external view from the local AR user using an HMD. They showed that all the views were useful in reducing the reliance on verbal instructions during the study.

While several studies have investigated the significance of key design features in collaborative AR, there has been limited exploration of how various design factors could impact users' collaborative and competitive activities in remote MAR contexts.

\section{Implementation of the AR Application}
We designed a mobile augmented reality application with all five techniques, allowing users to remotely interact with the virtual objects that were overlaid on top of the physical world. We wanted to focus on collaborative and competitive practices based on different design attributes rather than the complexity of models. \update{We thus used two virtual models/objects: (i) a 3D Rubik's cube with 8 colors of triangular and square faces and (ii) a 3D city with 85 cars of 10 different colors placed in random locations in the model}. We used a marker-based solution to design the AR application, where users can hold their smartphone at the marker to render the virtual objects on their mobile device screen.

We developed the prototype \update{application\footnote{https://github.com/NelushN/RemoteMAR}} using Unity with C\#. The virtual objects were overlaid on AR markers using Vuforia \cite{Vuforia2024}. We used Photon Unity Networking \cite{photonpun22023} to enable remote collaborations. In the application, we allow the first user who joins the application early to create a virtual room, and all the other users can join the same room (with the room name) to interact with the same 3D model. Each user is required to have the same AR markers to see the models. In addition, our application allows users to navigate to the main menu from the study screen and options to seek ``help'' from the user interface with ``?'' icon (Figure \ref{fig:overview3}). The ``help'' menu is used to access help related to the AR applications, such as creating a new room for collaboration, manipulating 3D content, and adding/removing annotations.

\section{Design Space Exploration}
Previous work demonstrates how different collaboration styles, strategies, and device configurations affect co-located collaboration \cite{10.1145/3313831.3376541,Wells2022:Co-Located_Collaboration}. 
Based on the prior work, we considered the following factors in designing solutions for remote collaborative and competitive tasks: 

\subsection{Content Viewing Angle}
The content viewing angle refers to the angle at which the AR content is captured and displayed on users' devices. Generally, on a solo exploration, the content viewing angle is the viewing angle at which the user holds their smartphone to the virtual object. However, when we have more than one user, we can have the following two styles: 

\textbf{Same Viewing Angle:}
In this style, both users view the same content from a specific angle, regardless of their devices' orientation relative to the virtual objects. The viewing angle is determined by the current device angle of one user, typically the user who joined the virtual room first. If the other user (i.e., User 2) wishes to change the viewing angle to their own, they can tap their screen to take ownership of the viewing angle. The other user then receives a prompt indicating that the viewing angle will switch shortly. One advantage of this style is that it allows both users to see the content from the same perspective, thus promoting consistency in their shared viewing experience. However, a disadvantage is that it may restrict users' flexibility in viewing the content from a different angle based on their preferences.
 
\textbf{Different Viewing Angle:}
Each user has their own view, determined by the angle at which they hold the device to point to the virtual content. One advantage of using this style is that each user can work independently on the shared 3D content. However, this model does not provide information on exactly where the other user is looking, thus limiting awareness of other users' activities in remote tasks.  

\subsection{Model Manipulation}
Since both users interact with the same virtual content, we have considered the impact of having different styles of synchronization of model manipulations on users' performance. We allow users to two-finger pinch-in and pinch-out to zoom in and zoom out the models and one-finger movements for translations. During implementation, we observed that rotating 3D objects around the 3 axes (i.e., x, y, and z) with 2D touch input was uncomfortable. Therefore, we opted for the rotation of the AR marker and the rotation of the mobile phone around the objects, instead of using touchscreen rotations. We considered the following two styles of model manipulations:

\textbf{Synchronous:}
In this manipulation style, both users observe the same manipulation outcome synchronously, regardless of who performs the action. For example, if User 1 pinches content to zoom in, User 2 also sees the zoomed-in version of the 3D model. We anticipate that this style of model manipulation would foster a shared understanding between users, as they both observe the effects of manipulation actions simultaneously. However, a potential disadvantage may lead to conflicts or confusion if one user has different preferences regarding manipulating the content.

\textbf{Asynchronous:}
In asynchronous manipulation, each user has their own model and independently performs manipulations that do not affect the other user's view or model. This style of manipulation allows users to explore and manipulate content without impacting each other's actions. However, a disadvantage is that using this style of manipulation may lead to miscommunication between users if they are not aware of each other's actions in real-time.

\subsection{Views}
View refers to having the entire screen show the virtual content or having an additional small view on top of the \update{personal} view showing what the other user is doing. This design is inspired by the ``speaker view'' opinion for video conferencing software such as Zoom \cite{zoom2023}, which allows users to see the speaker's content with a large window while showing the other user's view with small windows. Here are more details on these two options:

\begin{figure}[h]
    \centering
    \includegraphics[width=1.0\columnwidth]{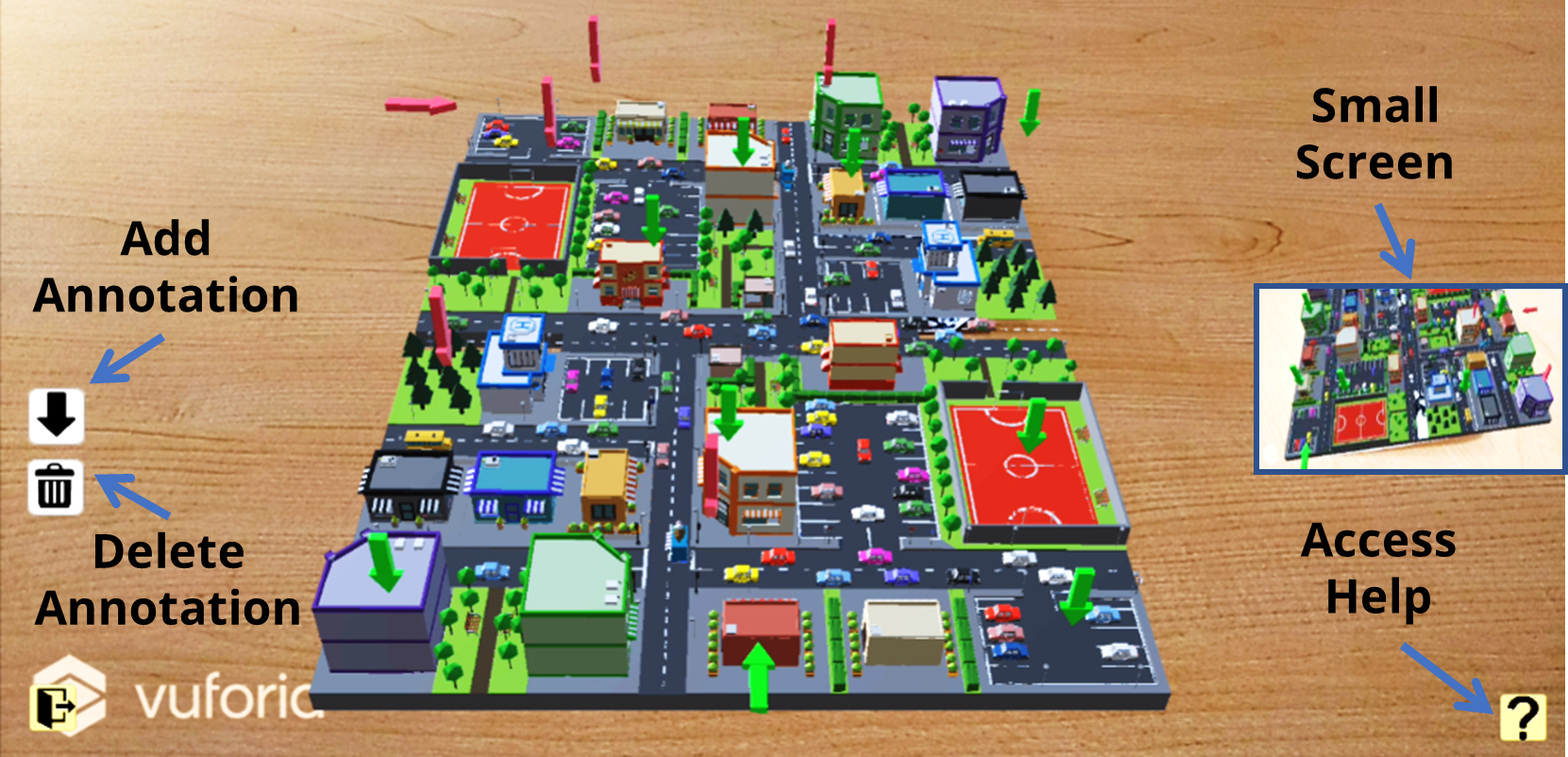}
    \caption{\update{Combined view of the Personal} View and Small \update{collaborator screen} View (marked with blue colored border)}
    \label{fig:overview3}
\end{figure} 

\textbf{\update{Personal} View:}
The \update{personal} view takes up the available space on their smartphones to show the AR scene. In addition, this view contains information such as the player/user ID and the name of the room they connected to. One advantage of the \update{personal} view is that it allows users to fully engage with their own AR content without distractions. However, this view may limit a user's awareness of other user's actions, potentially hindering remote collaboration.

\textbf{\update{Combined} Views:}
In this combination, the users see their \update{personal} view, along with the other user's interface within a small screen on the middle-right edge of the screen (Figure \ref{fig:overview3}). This feature lets users get an idea of each other's activities when performing the tasks, facilitating enhanced collaboration and coordination in remote tasks. A disadvantage of this view is that it consumes valuable \update{personal } screen space on small devices like smartphones.

\subsection{Additional Features}
Our application also lets users add and delete 3D arrow annotations to 3D models. We anticipate that the feature would be invaluable when participants are working remotely on complex objects, allowing them to pinpoint specific locations in the 3D environment and maintain spatial awareness of each other's activities. In our application, User 1's annotations are green, and User 2's are magenta, allowing participants to distinguish between their own and others' annotations easily.

\subsection{Techniques}

Based on the above factors, we designed a set of techniques as shown in Table \ref{tab:design_factors}. Although the combination of all these factors could potentially create a large number of techniques, we only focus on the techniques that are practical for remote collaborative and competitive tasks. For instance, using the ``Same Angle - Asynchronous Model Manipulations - \update{Personal} Screen'' combination was found to be ineffective, as participants continually needed to negotiate the ownership of control due to the change in angle with their phone orientation. This setup negated the benefits of asynchronous model manipulations. Likewise, implementing the Same Angle approach for techniques with a small screen proved to be unnecessary, as participants could easily view each other's point of view (POV) from their own dedicated, separate screens. \update{Therefore the authors considered all the possible combinations of different viewing attributes and chose the following five techniques to explore further.}

\begin{table}[h]
    \caption{Overview of Proposed Techniques}

\footnotesize
\centering
\begin{tabular}{ccccc}
\hline
\textbf{Technique} & \textbf{Viewing Angle} & \textbf{Manipulation} & \textbf{Views} \\
\hline
 \update{SSP}  & Same    & Synchronous & \update{Personal} \\
 \hline
  \update{DSP}  & Different   & Synchronous & \update{Personal}\\
  \hline
 \update{DAP} & Different    & Asynchronous & \update{Personal}\\
  \hline
 \update{DSC}  & Different    & Synchronous & \update{Combined}\\
  \hline
 \update{DAC}  & Different    & Asynchronous & \update{Combined}\\
\hline
\end{tabular}

    \label{tab:design_factors}
\end{table}

\begin{figure}[htp]
    \centering
    \includegraphics[width=1.0\columnwidth]{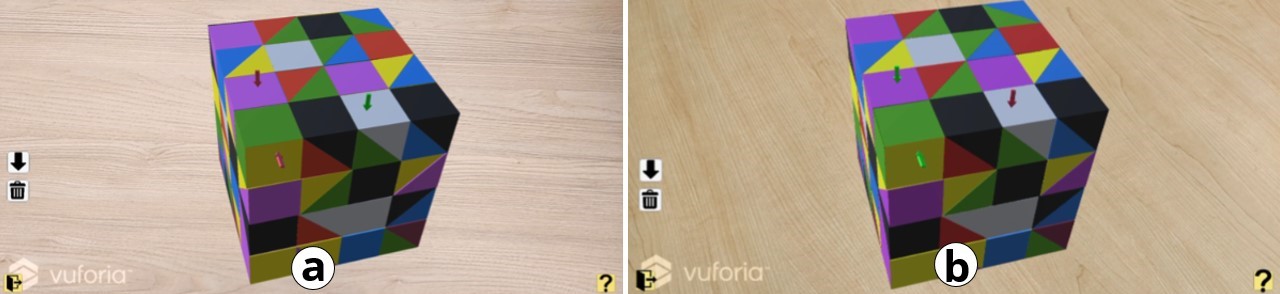}
    \caption{\update{SSP}: Same Angle - Synchronous Model - \update{Personal} View. \update{Views of (a) User 1 and (b) User 2. Both users see the cube with the same viewing angle, model size and rotation.}}
    \label{fig:ssl}
\end{figure}

\textbf{\update{SSP} (Same Angle - Synchronous Model - \update{Personal} View):} In this technique, both users can view the virtual contents from a specific angle, which is the viewing angle of the user who currently has the ownership of controlling the model. Once the owner manipulates the content, the other user also observes the same model. Note that this technique does not offer any additional screen (i.e., small screen) as both users can see the same content from their personal view. Figure \ref{fig:ssl}a and Figure \ref{fig:ssl}b show how User 1 and User 2 see the cube with the same POV (i.e., the same viewing angle) with the same model size and rotation. This feature allows the user to understand how exactly the other user is looking at the virtual object. As mentioned before, the user who wants to take control of the view can tap on the touchscreen and take control. The major disadvantage of this technique is that  only one person can control the virtual content at one time - thus limiting collaboration possibilities. 

\begin{figure}[h!]
    \centering
    \includegraphics[width=1.0\columnwidth]{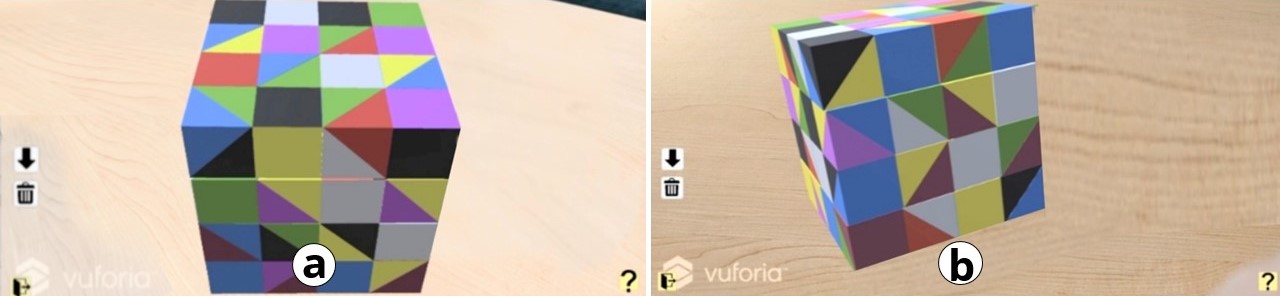}
    \caption{\update{DSP}: Different Angle - Synchronous Model - \update{Personal} view. \update{Views of (a) User 1 and (b) User 2. Users see the cube from different viewing angles based on their perspective, but have the same model size and rotation.}}
    \label{fig:dsl}
\end{figure}

\textbf{\update{DSP} (Different Angle - Synchronous Model - \update{Personal} view):} The \update{DSP} technique allows users to manipulate models (i.e., virtual contents) without sharing the same viewing angle. Each user can have their own viewing angle, determined by how they hold their smartphone. However, model manipulations are synchronized between users in their \update{personal} view (Figure \ref{fig:dsl}). Figure \ref{fig:dsl}a is an example where User 1 views the object from his own perspective, while Figure \ref{fig:dsl}b shows User 2 viewing the object from his own POV. As can be seen, both see the same scaled object since model manipulations are synchronized. One advantage of this technique is that collaborators can inspect the object from their own perspective without needing to align with other's viewing angles. However, as each user's viewing angle may differ, they might face difficulties in creating a spatial map of what the other users are doing during remote tasks.

\begin{figure}[h!]
    \centering
    \includegraphics[width=1.0\columnwidth]{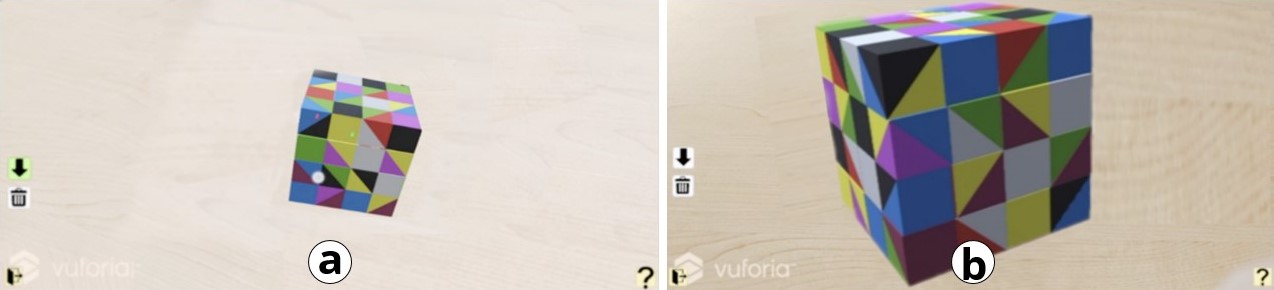}
    \caption{\update{DAP}: Different Angle - Asynchronous Model - \update{Personal} view. \update{Views of (a) User 1 and (b) User 2. Users see the cube differently - in size and rotation - based on their respective viewing angle.}}
    \label{fig:dal}
\end{figure}

\textbf{\update{DAP} (Different Angle - Asynchronous Model - \update{Personal} view):} With this technique, each user can manipulate their own 3D content without observing others' actions on their models. It does not allow users to share the same viewing angles as \update{SSP} does, nor does it allow users to leverage synchronized manipulations. Instead, users need to rely on annotations to understand each other's actions. Figure \ref{fig:dal}a shows an example where User 1 adds an annotation while User 2 scales and rotates the model (Figure \ref{fig:dal}b). One advantage of this technique is that it allows users to have full control over the virtual content giving them the flexibility to manipulate the model and interact with it independently. However, it can be disadvantageous when all collaborators need to work on the same object but cannot share the same viewing angle.

\begin{figure}[h]
    \centering
    \includegraphics[width=1.0\columnwidth]{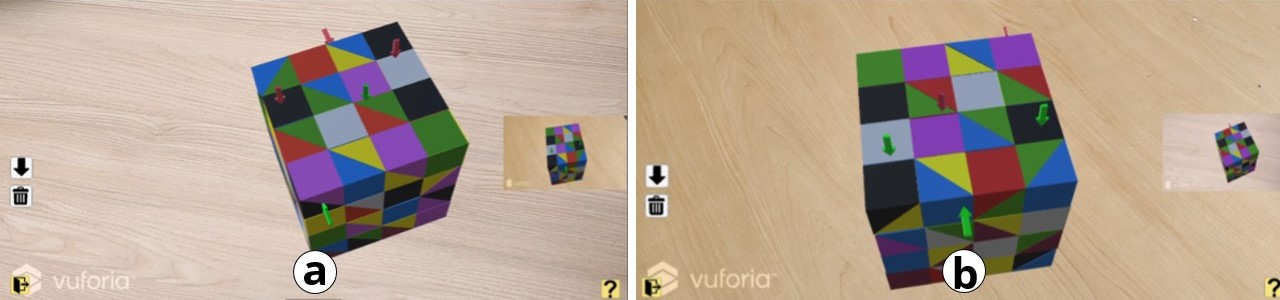}
    \caption{DSC: Different Angle - Synchronous Model - \update{Combined views}. \update{Users see the cube with different viewing angles; however, the model is synchronized in size and rotation. The small screen allows (a) User 1 to see User 2's personal view and (b) user 2 can see User 1's personal view.}}
    \label{fig:dsb}
\end{figure}

\textbf{DSC (Different Angle - Synchronous Model - \update{Combined views}):} One of the key features of this technique is that it allows both users to observe  on a small screen what the other user is doing. Figure \ref{fig:dsb}a shows an example of the technique where User 1 can see what User 2 is doing on the small screen. The \update{\update{Personal}} view displays the model from User 1's perspective, and the small screen displays the model from the perspective of User 2. The users do not share the same viewing angle but the manipulation of the model and the annotations are synchronized. 

\begin{figure}[!h]
    \centering
    \includegraphics[width=1.0\columnwidth]{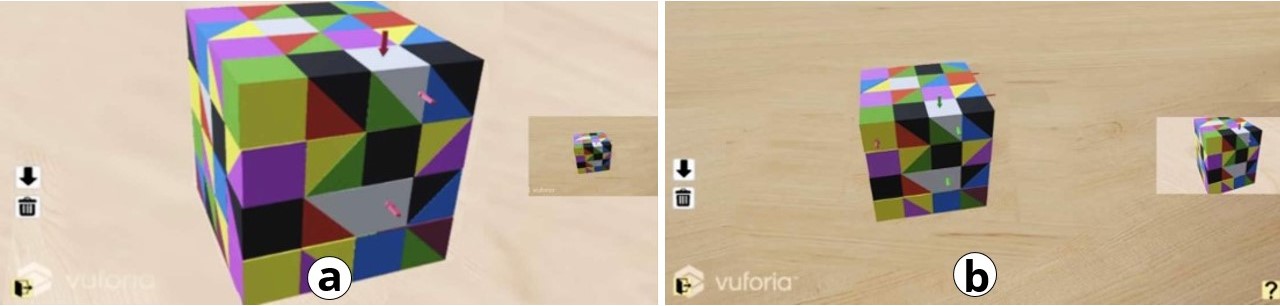}
    \caption{DAC: Different Angle - Asynchronous Model - \update{Combined views}. \update{Users see the cube differently - from different viewing angles with different cube sizes and rotation angles. Similar to \update{DSC}, a small screen is available to (a) User 1 and  (b) User 2 to see each other's personal views.}}
    \label{fig:dab}
\end{figure}

\textbf{DAC (Different Angle - Asynchronous Model - \update{Combined views}):} This technique has very similar features to the DSC technique despite the fact that the manipulations are not synchronized. Figure \ref{fig:dab} shows the technique where User 1 is scaling the model, and it is not synchronized to User 2. However, User 2 can see that User 1 is scaling his model from the small window.

\section{User Study}

The goal of this user study is to explore how different techniques can affect collaborative and competitive tasks when performed in remote settings. We selected collaborative tasks based on the groundwork laid by Wells et al. \cite{10.1145/3313831.3376541}. Recognizing a gap in the existing research specifically aimed at design attributes for remote collaborative and competitive tasks, we focused on evaluating the impact of our designed techniques on task completion efficiency and effectiveness.

\subsection{Participants, Tasks, and Procedure}
We recruited 20 participants (11 male and 9 female), aged between 21 and 34 years (mean: 26.65 years, SD: 4.21 years), using snowball sampling. The participants were paired to represent (i) one local user, i.e., User 1, and (ii) a remote user, i.e., User 2. There were no requirements for the study other than the participants not being color blind as tasks required to identify colors. Twelve participants had prior experience using AR applications and reported using AR applications once in the past 12 months. However, all participants had prior experience using mobile devices. Each participant was compensated with CAD \$15.

\update{Participants were asked to perform counting tasks collaboratively and competitively for each technique while inspecting each virtual model (3D Rubik's cube and 3D city models).} The participants were given a consent form before the study. Once both participants arrived at the study location (i.e., a lab at a university), we collected the signed consent forms. We then gathered their demographic information, including age, gender, and experience with using AR technology and touchscreen handheld mobile devices. Participants were next given the study details, such as how the 3D models can be seen on the screen, how to manipulate the models, as well as describing the tasks they will be required to perform (counting specific colored shapes or colored cars dependent on the model). 
The participants were then invited into two different rooms for the study. Each of them was provided with a smartphone (i.e., either a Google Pixel 3a mobile phone or a One Plus 7T mobile) with the AR application installed for the study. Though the two rooms were completely isolated, both participants were connected via a Zoom call to facilitate voice communication. They were given a set of practice trials with both models for each types of task (collaborative and competitive) to familiarize themselves with the application. They were instructed to move around the image marker placed on a table or move/rotate the marker image printed on paper to manipulate the 3D objects. They were encouraged to discuss strategies that they wanted to use during the study session - e.g., placing arrows as annotations to show which item they have already counted - specifically for collaborative tasks. 

\update{Each task began with an experimenter asking a question. For the Rubik’s Cube, questions focused on the shape of the faces, i.e., triangular or square, and their colors, such as \textit{``Count the number of X-colored triangular faces on the Rubik’s Cube''}. In contrast, for the city model, questions were centered on car colors such as \textit{``Count the number of Y-colored cars in the city''} where X or Y represents a color. For instance, in Figure \ref{fig:dab}, participants were asked to count the grey-colored triangles on the Rubik’s Cube. In collaborative tasks, both participants were required to reach a consensus on their answers before communicating the result to the experimenter. Furthermore, participants were permitted to add/remove arrows to the 3D models to indicate those already visited - a task concluded upon any participant providing their agreed upon answer. For competitive tasks, participants individually completed counting tasks and reported their results to the experimenter. A task concluded when both participants had responded, and the experimenter verified the correctness before proceeding to the next task.}
The experimenter assisted participants in loading the correct model and appropriate technique for each task. Each session of the user study was video recorded for a detailed analysis of each participant's collaborative and competitive behavior patterns in a remote context. 

Techniques were counterbalanced across each group using a Latin square to account for learning effects during the study. For each techniques, participants completed the counting tasks for the cube model, first collaboratively then individually, before proceeding to counting tasks in the city model in the same sequence. Thus, each group underwent 5 \textit{Technique} \update{(SSP, DSP, DAP, DSC, DAC)} $\times$ 2 \textit{Model} (cube and city) $\times$ \textit{Style} (competitive and collaborative) = 20 unique questions or trials. The study session lasted approximately 60 minutes.

\subsection{Measures}
We recorded task completion time which was calculated as the duration in seconds between the time when participants started a trial and the time when they answered the question. 
We used NASA-TLX \cite{NASA-TLX} to measure the perceived workload for performing both collaborative and competitive tasks for each technique. We further used SUS \cite{SUS} to measure the usability of each technique. We also asked the participants to provide ratings on their overall preferences for each technique on a 7-point Likert scale.

We also examined all the video footage recorded during the study. Following the prior works \cite{Zagermann2016:Tablets_meet_Tabletops, 10.1145/3313831.3376541, Wells2022:Co-Located_Collaboration}, two researchers coded the videos categorizing participants' behavior and strategies into three main categories: (i) Communication, (ii) Annotation, and (iii) Manipulation. We also observed their movement (i.e., standing, seated) and device orientation (i.e., portrait, landscape) during each task across the techniques. The videos were coded using NVivo software \cite{NVivo}. 

\section{Results}

\subsection{Task Completion Time}
To analyze task completion times, we used two-way repeated-measures Analysis of variance (ANOVA) with post-hoc pairwise comparisons (Bonferroni adjusted).

\begin{figure}[h]
    \centering
    \includegraphics[width=1.0\columnwidth]{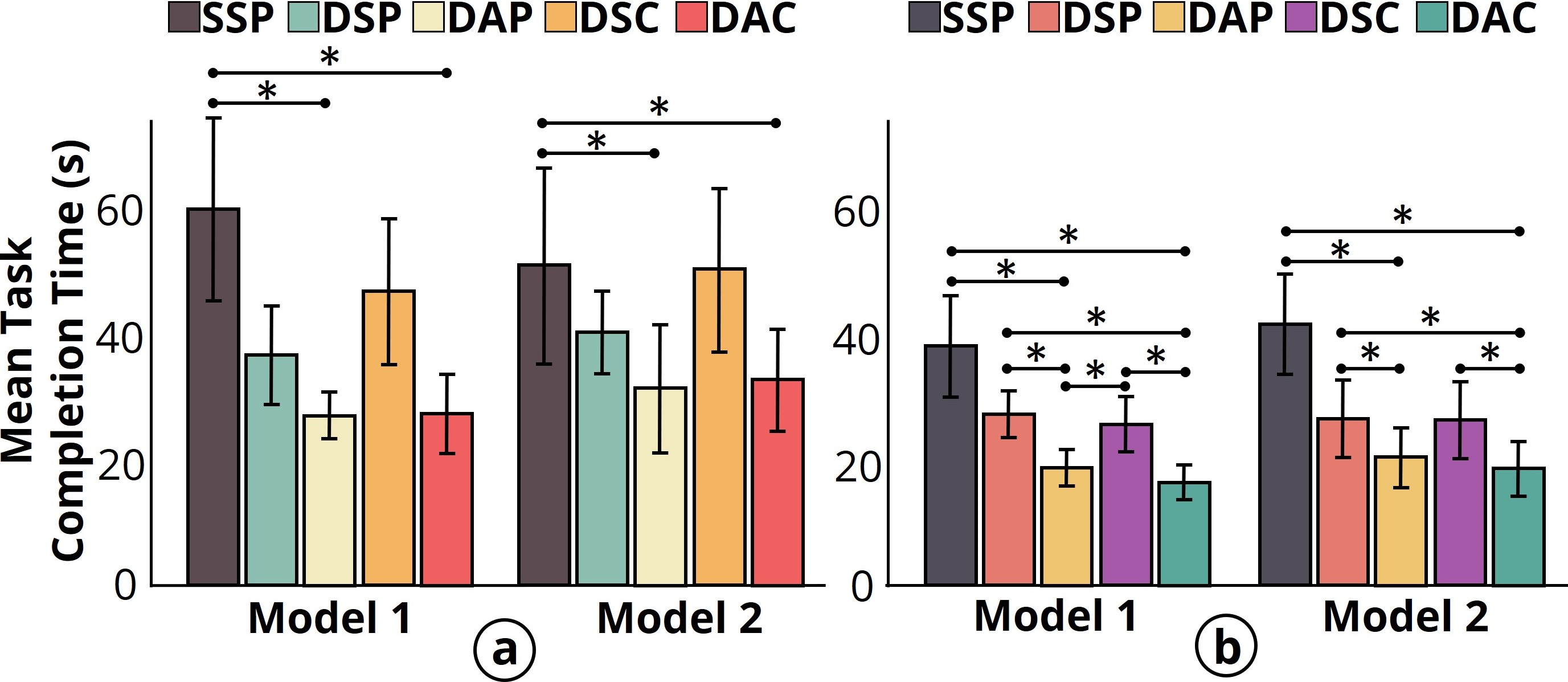}
    \caption{Mean task completion time: (a) collaborative and (b) competitive tasks by each technique across both models. Statistical significances are marked with $\ast$ [Error bars: 95\% CI].  }
    \label{fig:collaborative_time}
\end{figure}

\textbf{Collaborative Tasks:}
Results from repeated-measure ANOVA showed \textit{Model} has no main effect on task completion times ($F_{(1,9)} = 0.17,$ $p = 0.69,$ $\eta_P^2 = 0.02$). 
However, we found significant main effect of \textit{Techniques} ($F_{(4,36)} = 7.33, p < 0.001,$ $\eta_P^2 = 0.45$) on task completion times. Pairwise comparisons showed that participants were significantly faster with \update{DAP} ($M = 28.80s$) and \update{DAC} ($M = 29.80s$) compared to \update{SSP} ($M = 55.10s$). No significant differences were noticed between other pairs. No interaction between \textit{Model} and \textit{Technique} was found. Figure \ref{fig:collaborative_time}(a) shows the mean task completion time for collaborative tasks taken by each technique across both of the models.

\textbf{Competitive Tasks:}
Results showed that \textit{Model} has no main effect on task completion times ($F_{(1,19)} = 0.56,$ $p = 0.47,$ $\eta_P^2 = 0.03$). 
However, we noticed significant main effect of \textit{Techniques} ($F_{(4,76)} = 35.53, p < 0.001,$ $\eta_P^2 = 0.65$) on task completion times. Participants were significantly faster with \update{DAP} ($M = 19.33s$) compared to \update{SSP} ($M = 39.45s$), \update{DSP} ($M = 26.55s$) and \update{DSC} ($M = 25.48s$). Similarly, \update{DAC} ($M = 17.43s$) was significantly faster than \update{SSP}, \update{DSP} and \update{DSC}. Although the mean task completion time for \update{DAC} was lower than \update{DAP}, no significant difference ($p = 1.00$) was found between them. Moreover, no interaction between \textit{Model} and \textit{Technique} was noticed. Figure \ref{fig:collaborative_time}(b) shows the mean task completion time for competitive tasks taken by each technique across both of the models.

\begin{table*}[!h]
    \caption{Perceived workload (NASA-TLX) for collaborative and competitive tasks in all sub-scales between the techniques where $p < 0.005$.}
    \resizebox{\textwidth}{!}{%
\begin{tabular}{ |c|c|c|c|c|c|c|c|}
 \hline
 & \multirow{8}{*}{\rotatebox{90}{Collaborative}} & Mental Demand & Physical Demand & Temporal Demand & Performance & Effort & Frustration\\
 \cline{1-1} \cline{3-8}
 \update{DAP} $\leftrightarrow$ \update{SSP} & & Z = -3.98, $p <$ 0.001 & Z = -3.98, $p <$ 0.001 & Z = -3.91, $p <$ 0.001 & Z = -3.93, $p <$ 0.001 & Z = -3.98, $p <$ 0.001 & Z = -4.01, $p <$ 0.001\\
 \cline{1-1} \cline{3-8}
 \update{DAP} $\leftrightarrow$ \update{DSP} & & Z = -3.57, $p <$ 0.001 & Z = -3.85, $p <$ 0.001 & Z = -3.84, $p <$ 0.001 & Z = -3.42, $p <$ 0.001 & Z = -3.83, $p <$ 0.001 & Z = -3.42, $p <$ 0.001\\
 \cline{1-1} \cline{3-8}
 \update{DAP} $\leftrightarrow$ \update{DSC} & & Z = -3.04, $p$ = 0.002 & Z = -3.64, $p <$ 0.001 & Z = -3.51, $p <$ 0.001 & Z = -2.94, $p$ = 0.003 & Z = -2.99, $p$ = 0.003 & Z = -3.00, $p$ = 0.003\\
 \cline{1-1} \cline{3-8}
 \update{DAP} $\leftrightarrow$ \update{DAC} & & Z = -3.94, $p <$ 0.001 & Z = -3.31, $p <$ 0.001 & Z = -2.93, $p$ = 0.003 & Z = -3.93, $p <$ 0.001 & Z = -3.35, $p <$ 0.001 & Z = -3.49, $p <$ 0.001\\
 \cline{1-1} \cline{3-8}
 \update{DAC} $\leftrightarrow$ \update{SSP} & & Z = -3.96, $p <$ 0.001 & Z = -3.95, $p <$ 0.001 & Z = -3.85, $p <$ 0.001 & Z = -3.97, $p <$ 0.001 & Z = -3.95, $p <$ 0.001 & Z = -3.96, $p <$ 0.001\\
 \cline{1-1} \cline{3-8}
 \update{DAC} $\leftrightarrow$ \update{DSP} & & Z = -3.87, $p <$ 0.001 & Z = -3.84, $p <$ 0.001 & Z = -3.90, $p <$ 0.001 & Z = -3.94, $p <$ 0.001 & Z = -3.97, $p <$ 0.001 & Z = -3.97, $p <$ 0.001\\
 \cline{1-1} \cline{3-8}
 \update{DAC} $\leftrightarrow$ \update{DSC} & & Z = -3.69, $p <$ 0.001 & Z = -3.87, $p <$ 0.001 & Z = -3.87, $p <$ 0.001 & Z = -3.95, $p <$ 0.001 & Z = -3.57, $p <$ 0.001 & Z = -3.62, $p <$ 0.001\\
 \cline{1-1} \cline{2-2} 
 \cline{3-8}
 
 \update{DAP} $\leftrightarrow$ \update{SSP} & \multirow{5}{*}{\rotatebox{90}{\hspace{-.5em}Competitive}} &
  Z = -3.97, $p <$ 0.001 & 
  Z = -3.87, $p <$ 0.001 & 
  Z = -3.96, $p <$ 0.001 & 
  Z = -3.84, $p <$ 0.001 & 
  Z = -3.98, $p <$ 0.001 & 
  Z = -3.96, $p <$ 0.001\\
 \cline{1-1} \cline{3-8}
 \update{DAP} $\leftrightarrow$ \update{DSP} & &
  Z = -4.00, $p <$ 0.001 & 
  Z = -3.83, $p <$ 0.001 & 
  Z = -3.68, $p <$ 0.001 & 
  Z = -3.54, $p <$ 0.001 & 
  Z = -3.91, $p <$ 0.001 & 
  Z = -2.85, $p$ = 0.004\\
 \cline{1-1} \cline{3-8}
 \update{DAC} $\leftrightarrow$ \update{SSP} & &
  Z = -3.98, $p <$ 0.001 & 
  Z = -3.95, $p <$ 0.001 & 
  Z = -3.96, $p <$ 0.001 & 
  Z = -3.86, $p <$ 0.001 & 
  Z = -3.96, $p <$ 0.001 & 
  Z = -3.96, $p <$ 0.001\\
 \cline{1-1} \cline{3-8}
 \update{DAC} $\leftrightarrow$ \update{DSP} & &
  Z = -3.88, $p <$ 0.001 & 
  Z = -3.87, $p <$ 0.001 & 
  Z = -3.96, $p <$ 0.001 & 
  Z = -3.77, $p <$ 0.001 & 
  Z = -3.87, $p <$ 0.001 & 
  Z = -3.77, $p <$ 0.001\\
 \cline{1-1} \cline{3-8}
 \update{DAC} $\leftrightarrow$ \update{DSC} & & 
  Z = -3.72, $p <$ 0.001 & 
  Z = -3.82, $p <$ 0.001 & 
  Z = -3.98, $p <$ 0.001 & 
  Z = -3.74, $p <$ 0.001 & 
  Z = -3.34, $p <$ 0.001 & 
  Z = -3.56, $p <$ 0.001\\
 \hline
\end{tabular}%
}

    \label{tab:collab_comp_workload}
\end{table*}

\subsection{Workload}
We used Friedman tests with Wilcoxon signed rank tests (for post-hoc pairwise comparisons) to analyze the workload data. Post-hoc pairwise comparisons were Bonferroni adjusted ($\alpha$-level 0.05 to 0.005).

\begin{figure}[!h]
    \centering
    \includegraphics[width=1.0\columnwidth]{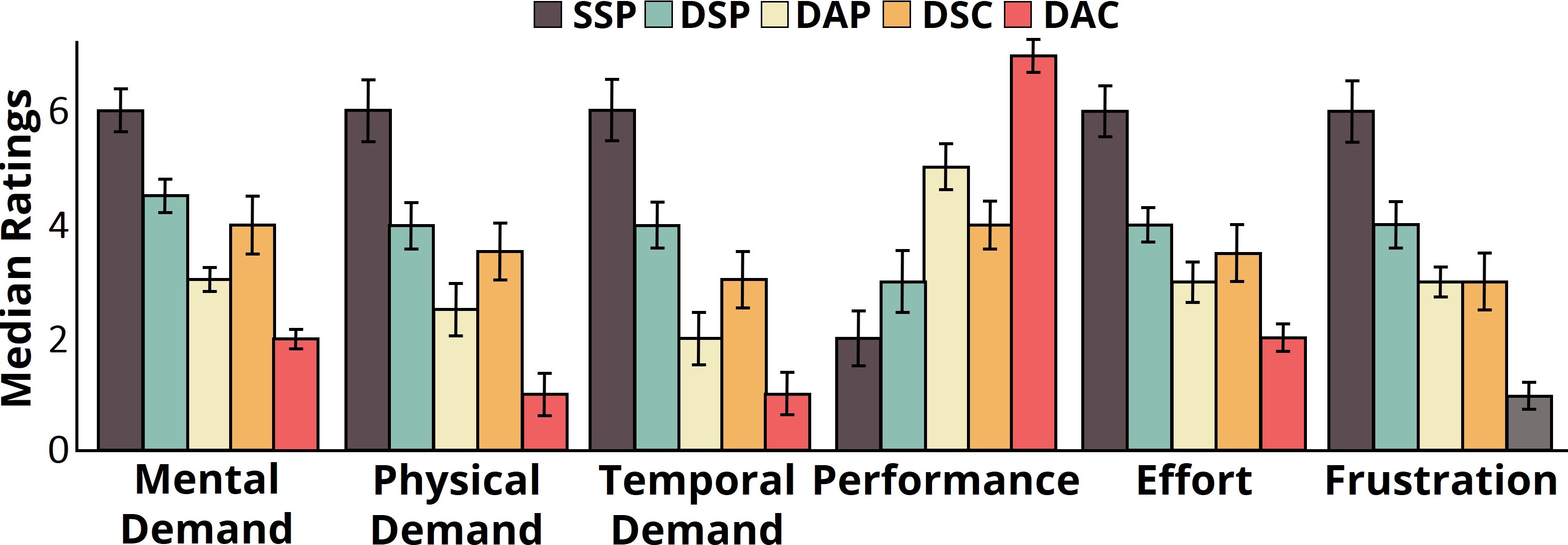}
    \caption{The median ratings for NASA-TLX categories across all the techniques for performing collaborative tasks  [Error bars: 95\% CI]} 
    \label{fig:collaborative_workload}
\end{figure}

\textbf{Collaborative Tasks:}
 Results from the Friedman test showed statistically significant differences in all of the sub-scales: mental demand ($\chi^2(4, N = 20) = 70.73, p < 0.001$), physical demand ($\chi^2(4, N = 20) = 68.75, p < 0.001$), temporal demand ($\chi^2(4, N = 20) = 63.47, p < 0.001$), performance ($\chi^2(4, N = 20) = 68.45, p < 0.001$), effort ($\chi^2(4, N = 20) = 70.29, p < 0.001$), and frustration ($\chi^2(4, N = 20) = 65.53, p < 0.001$). From further pairwise comparisons, both \update{DAP} and \update{DAC} showed statistically significant differences ($p < 0.005$) with all other techniques. Moreover, \update{DAC} was rated most favorably and showed a statistically significant difference ($p < 0.005$) from \update{DAP}. \update{Table \ref{tab:collab_comp_workload}} summarises the results. Figure \ref{fig:collaborative_workload} shows participants' median ratings for NASA-TLX sub-scales across all the techniques. 

\begin{figure}[!h]
    \centering
    \includegraphics[width=1.0\columnwidth]{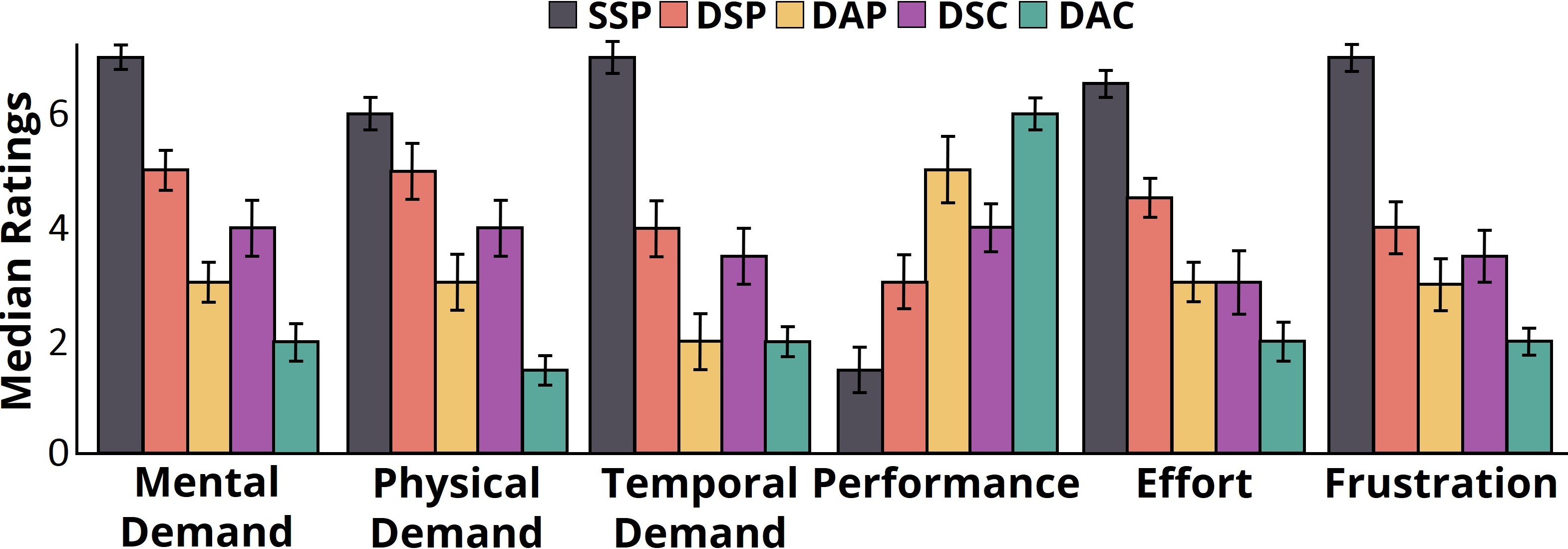}
    \caption{The median ratings for NASA-TLX categories across all the techniques for performing competitive tasks [Error bars: 95\% CI]}
    \label{fig:competitive_workload}
\end{figure}

\textbf{Competitive Tasks:}
Results from the Friedman test showed statistically significant differences in all of the NASA-TLX sub-scales: mental demand ($\chi^2(4, N = 20) = 67.54, p < 0.001$), physical demand ($\chi^2(4, N = 20) =67.25, p < 0.001$), temporal demand ($\chi^2(4, N = 20) = 70.22, p < 0.001$), performance ($\chi^2(4, N = 20) = 56.72, p < 0.001$), effort ($\chi^2(4, N = 20) = 67.25, p < 0.001$), and frustration ($\chi^2(4, N = 20) = 60.39, p < 0.001$). \update{DAC} showed statistically significant differences ($p < 0.005$) with all other techniques except \update{DAP}. \update{DAP} showed statistically significant differences ($p < 0.005$) with \update{SSP} and \update{DSP}. However, participants rated \update{DAC} most favorably in the NASA-TLX sub-scales. \update{Table \ref{tab:collab_comp_workload}} summarises the results. Figure \ref{fig:competitive_workload} shows participants' median ratings for NASA-TLX sub-scales. 

\subsection{Usability}
We used the following three statements from the SUS to analyze the usability of the techniques. S1: \textit{I thought the technique was easy to use}; S2: \textit{I found the various functions in this technique were well integrated}; and S3: \textit{I felt very confident using this technique}. 

\begin{figure}[h]
    \centering
    \includegraphics[width=1.0\columnwidth]{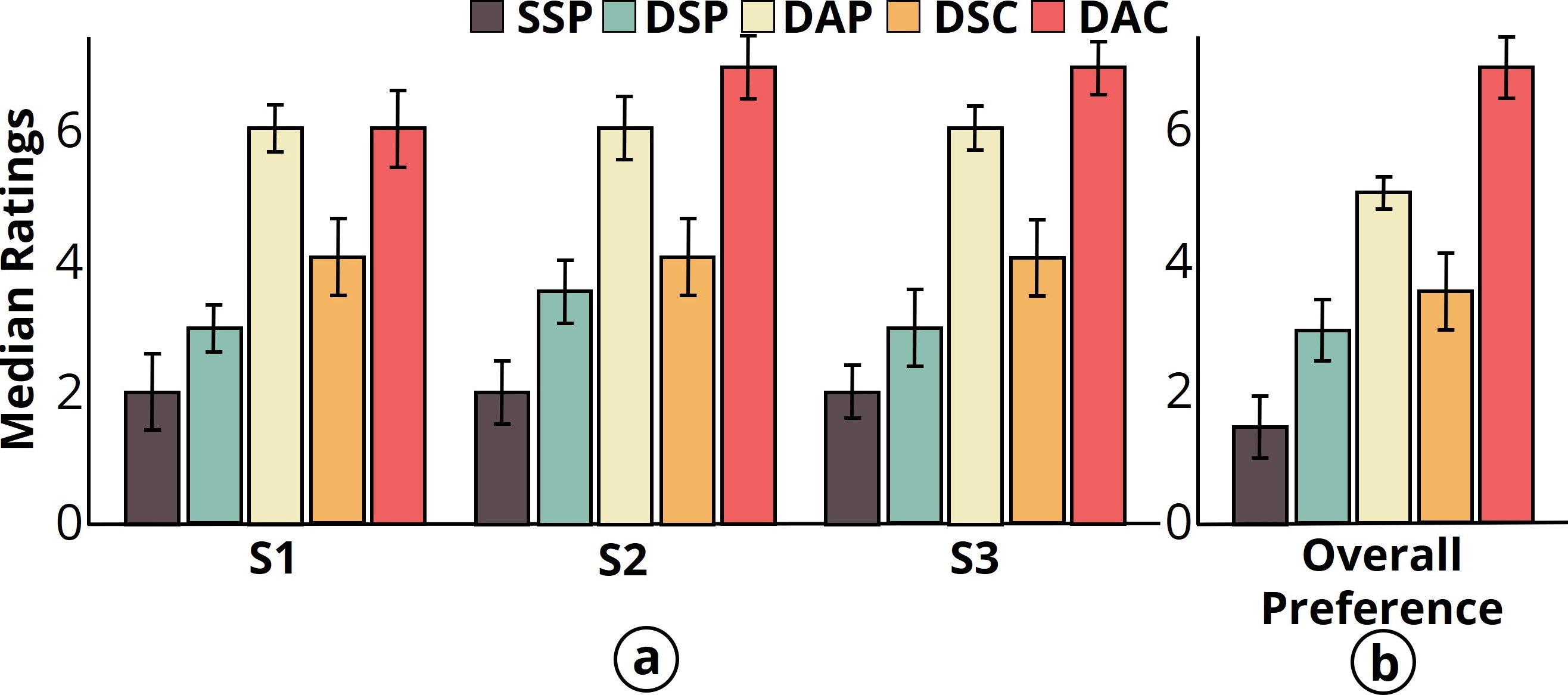}
    \caption{The median ratings on (a) usability and (b) overall preference across all the techniques for collaborative tasks [Error bars: 95\% CI]}
    \label{fig:collaborative_sus_pref}
\end{figure}

\textbf{Collaborative Tasks:}
Figure \ref{fig:collaborative_sus_pref} shows participants’ median responses to these three questions on a 7-point scale.
For S1, the Friedman test showed significant differences ($\chi^2(4, N = 20) = 58.51, p < 0.001$) among the techniques. \update{DAP} ($M = 5.80, SD = 0.83$) was significantly preferred over \update{SSP} ($M = 2.20, SD = 1.06,$ $Z = -3.94, p < 0.001$), \update{DSP} ($M = 3.50, SD = 0.95,$ $Z = -3.87, p < 0.001$) and \update{DSC} ($M = 4.10, SD = 1.21,$ $Z = -3.51, p < 0.001$). Similarly, \update{DAC} ($M = 6.15, SD = 1.09$) was significantly preferred over \update{SSP} ($Z = -3.88, p < 0.001$), \update{DSP} ($Z = -3.76, p < 0.001$) and \update{DSC} ($Z = -3.44, p < 0.001$). No significant differences ($Z = -1.45, p = 0.148$) were found between \update{DAC} and \update{DAP}.

Similarly, for S2, we observed a significant difference ($\chi^2(4, N = 20) = 59.05, p < 0.001$) among the techniques. Furthermore, \update{DAC} ($M = 6.40, SD = 0.99$) technique was significantly preferred over \update{SSP} ($M = 2.10, SD = 0.97,$ $Z = -3.85, p < 0.001$), \update{DSP} ($M = 3.50, SD = 1.05,$ $Z = -3.87, p < 0.001$) and \update{DSC} ($M = 4.00, SD = 1.21,$ $Z = -3.66, p < 0.001$). \update{DAP} ($M = 5.85, SD = 0.93$) was significantly preferred over \update{SSP} ($Z = -3.95, p < 0.001$), \update{DSP} ($Z = -3.82, p < 0.001$) and \update{DSC} ($Z = -3.37, p < 0.001$). There were no significant differences ($Z = -1.90, p = 0.058$) between \update{DAC} and \update{DAP}.

For S3, we also observed a significant difference ($\chi^2(4, N = 20) = 58.66, p < 0.001$) from the Friedman test. Pairwise tests demonstrated \update{DAC} ($M = 6.35, SD = 0.99$) technique being significantly preferred over \update{SSP} ($M = 2.10, SD = 0.85,$ $Z = -3.85, p < 0.001$), \update{DSP} ($M = 3.55, SD = 1.19,$ $Z = -3.75, p < 0.001$) and \update{DSC} ($M = 4.15, SD = 1.26,$ $Z = -3.62, p < 0.001$). \update{DAP} was significantly preferred than \update{SSP} ($Z = -3.96, p < 0.001$), \update{DSP} ($Z = -3.48, p < 0.001$) and \update{DSC} ($Z = -3.06, p = 0.002$). No significant differences ($Z = -1.90, p = 0.058$) were found between \update{DAC} and \update{DAP}, even though the \update{DAP} ($M = 5.70, SD = 0.86$) received a lower average rating.

\begin{figure}[!h]
    \centering
    \includegraphics[width=1.0\columnwidth]{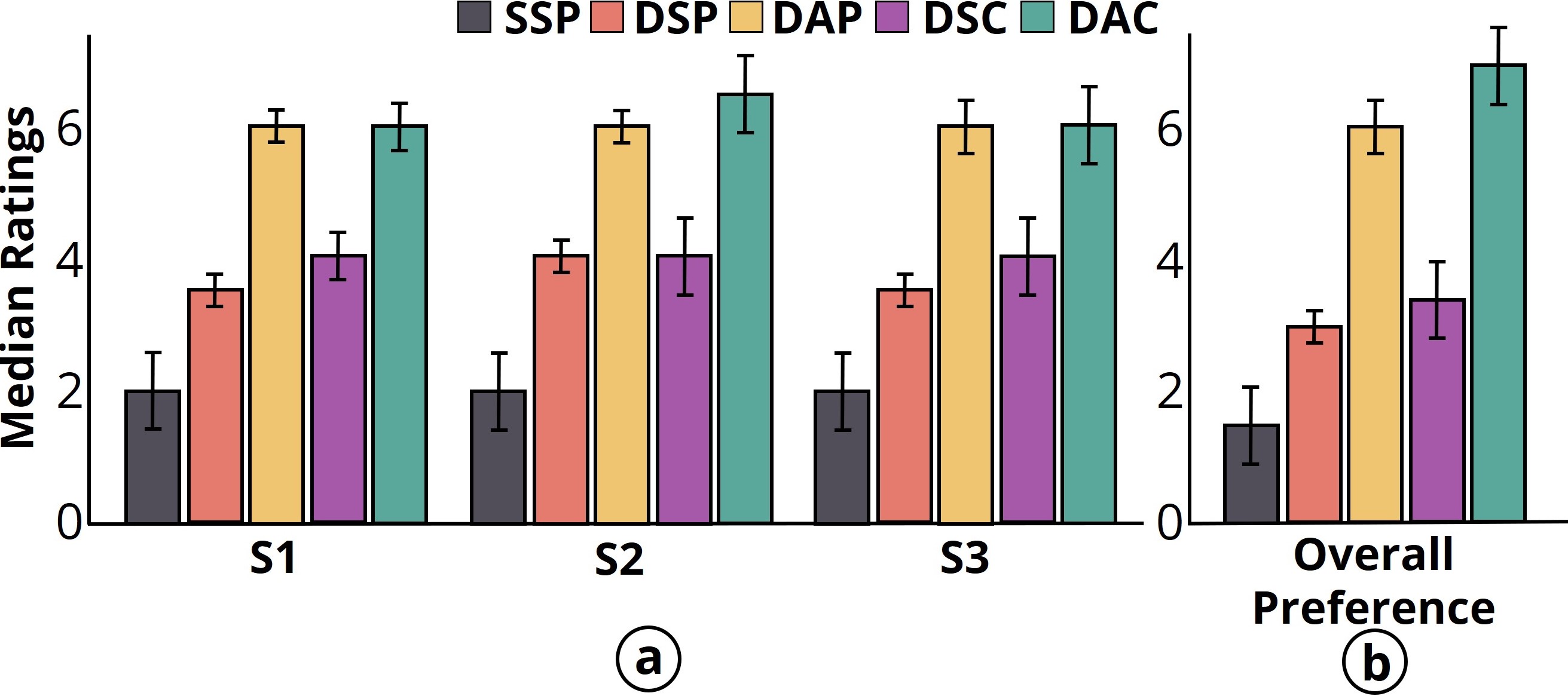}
    \caption{The median ratings on (a) usability and (b) overall preference across all the techniques for competitive tasks [Error bars: 95\% CI]}
    \label{fig:competitive_sus_pref}
\end{figure}

\textbf{Competitive Tasks:}
Figure \ref{fig:competitive_sus_pref} shows participants’ median responses to these three questions on a 7-point scale. 
For S1, the Friedman test revealed significant differences ($\chi^2(4, N = 20) = 54.46, p < 0.001$) among the techniques. Pairwise tests indicated that \update{DAP} ($M = 5.90, SD = 0.79$) was significantly preferred over \update{SSP}  ($M = 2.25, SD = 1.33,$ $Z = -3.85, p < 0.001$), \update{DSP} ($M = 3.55, SD = 0.76,$ $Z = -3.87, p < 0.001$) and \update{DSC} ($M = 4.25, SD = 1.29,$ $Z = -3.47, p < 0.001$). \update{DAC} ($M = 5.90, SD = 1.12$) was significantly preferred than \update{SSP} ($Z = -3.81, p < 0.001$), \update{DSP} ($Z = -3.66, p < 0.001$) and \update{DSC} ($Z = -3.10, p = 0.002$). However, no significant difference ($Z = -0.12, p = 0.915$) was found between \update{DAP} and \update{DAC}.

Similarly, for S2, significant difference ($\chi^2(4, N = 20) = 54.26, p < 0.001$) was noticed among the techniques. \update{DAC} ($M = 5.90, SD = 1.33$) was significantly preferred over \update{SSP} ($M = 2.10, SD = 1.25,$ $Z = -3.70, p < 0.001$), \update{DSP} ($M = 3.75, SD = 0.79,$ $Z = -3.42, p < 0.001$) and \update{DSC} ($M = 4.20, SD = 1.36,$ $Z = -3.02, p = 0.002$). \update{DAP} ($M = 5.80, SD = 0.70$) was preferred over \update{SSP} ($Z = -3.85, p < 0.001$), \update{DSP} ($Z = -3.87, p < 0.001$) and \update{DSC} ($Z = -3.21, p = 0.001$). No significant differences ($Z = -0.07, p = 0.942$) between \update{DAC} and \update{DAP}.

For S3, Friedman test showed a significant difference ($\chi^2(4, N = 20) = 54.40, p < 0.001$) across the techniques. Pairwise tests demonstrated \update{DAC} ($M = 5.95, SD = 1.15$) being significantly preferred over \update{SSP} ($M = 2.25, SD = 1.33,$ $Z = -3.69, p < 0.001$), \update{DSP} ($M = 3.55, SD = 0.76,$ $Z = -3.76, p < 0.001$) and \update{DSC} ($M = 4.30, SD = 1.26,$ $Z = -3.03, p = 0.002$). \update{DAP} ($M = 5.85, SD = 0.99$) was significantly preferred than \update{SSP} ($Z = -3.75, p < 0.001$), \update{DSP} ($Z = -3.86, p < 0.001$) and \update{DSC} ($Z = -3.24, p = 0.001$). No significant difference ($Z = -0.24, p = 0.81$) was found between \update{DAP} and \update{DAC}.

\subsection{Overall Preference}
We asked the participants to rate the techniques on a 7-point scale based on their overall preference.

\textbf{Collaborative Tasks:}
 Results from a Friedman test showed significant differences ($\chi^2(4, N = 20) = 66.54, p < 0.001$) among the techniques for performing collaborative tasks. Participants preferred \update{DAC} ($M = 6.70, SD = 0.92$) the most with significant differences than \update{SSP} ($M = 1.65, SD = 0.75,$ $Z = -3.90, p < 0.001$), \update{DSP} ($M = 3.15, SD = 0.88,$ $Z = -3.87, p < 0.001$), \update{DAP} ($M = 5.35, SD = 0.67,$ $Z = -3.26, p = 0.001$), and \update{DSC} ($M = 3.70, SD = 1.30,$ $Z = -3.87, p < 0.001$).

\textbf{Competitive Tasks:}
Results from another Friedman test showed significant differences ($\chi^2(4, N = 20) = 64.82, p < 0.001$) among the techniques for performing collaborative tasks. Pairwise comparison showed \update{DAC} ($M = 6.15, SD = 1.04$) had significant differences with \update{SSP} ($M = 1.75, SD = 1.02,$ $Z = -3.92, p < 0.001$), \update{DSP} ($M = 3.30, SD = 0.57,$ $Z = -3.86, p < 0.001$), and \update{DSC} ($M = 3.90, SD = 1.17,$ $Z = -3.75, p < 0.001$). \update{DAP} ($M = 5.85, SD = 0.75$) also showed significant differences with \update{SSP} ($Z = -3.95, p < 0.001$), \update{DSP} ($Z = -3.96, p < 0.001$), and \update{DSC} ($Z = -3.62, p < 0.001$). However, no significant difference ($Z = -0.81, p < 0.419$) was found between \update{DAP} and \update{DAC}.

\subsection{User behavior and strategies}
\begin{figure}[h]
    \centering
    \includegraphics[width=1.0\columnwidth]{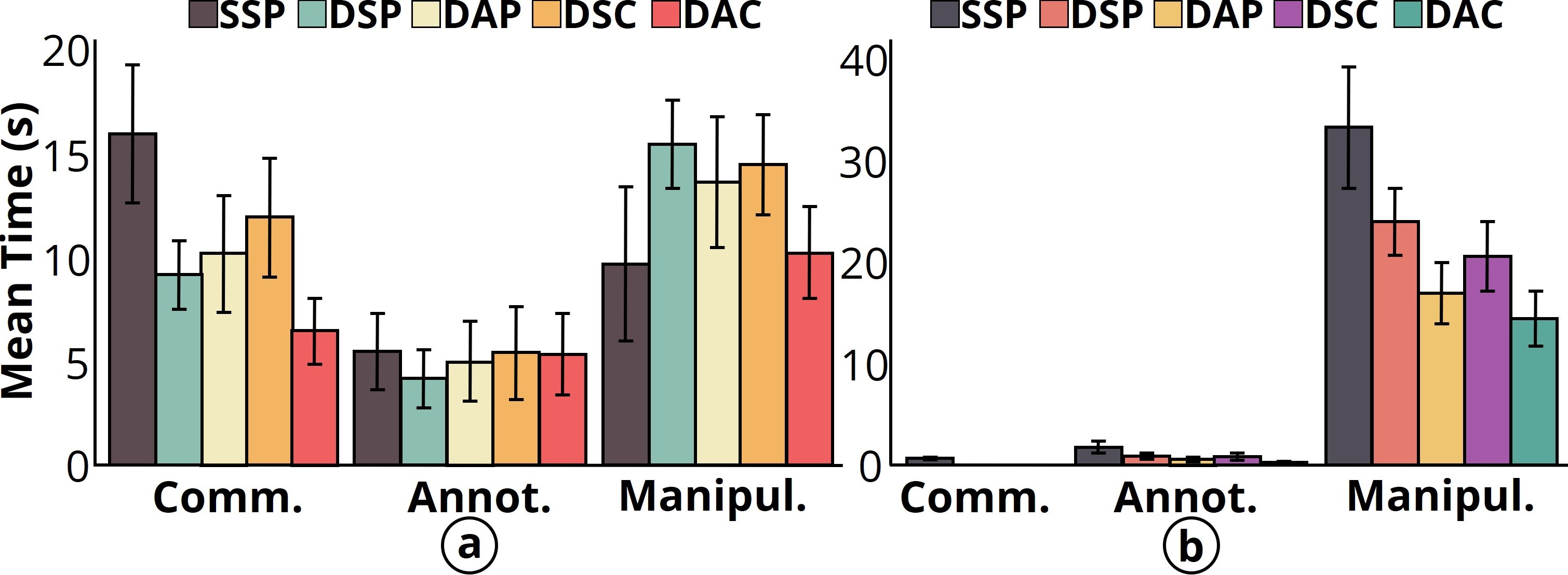}
    \caption{The mean times for communication, annotation, and manipulation during (a) collaborative \& (b) competitive tasks [Error bars: 95\% CI]}
    \label{fig:strategies}
\end{figure}

We examined the participants' behavior and strategies for different techniques across both collaborative and competitive tasks by analyzing the categories: (i) Communication, (ii) Annotation, and (iii) Manipulation derived from the coding scheme. Note that all these data were extracted from the videos captured during the study session. We used One-way ANOVA with Bonferroni corrected post-hoc tests to analyze these measures. Figure \ref{fig:strategies} shows the mean times taken for communication, annotation, and manipulation across all the techniques while performing collaborative and competitive tasks, respectively.
We did not analyze the participants' movement (i.e., standing, seated) and device orientation (i.e., portrait, landscape) as almost everyone remained seated, holding the smartphone in landscape orientation during the study. Only one participant from group 8 was standing/walking for the whole period while performing collaborative tasks in \update{DSC}. Participants only from group 5 held their phones in portrait orientation for some time while performing competitive tasks in \update{SSP} (75\% of the time), \update{DSC} (75\% of the time), and \update{DAC} (90\% of the time).  

\subsubsection{Communication }

\textbf{Collaborative Tasks:} One-way ANOVA showed that there is a significant difference in times spent for communication among the techniques while performing collaborative tasks ($F_{4, 195} = 7.76$, $p < 0.001$, $\eta^2 = 0.14$). Post-hoc tests showed that \update{DAC} ($M = 6.43s$) required significantly less communication time than \update{SSP} ($M = 15.83s, p < 0.001$) and \update{DSC} ($M = 11.95s, p = 0.021$). Although the differences were not statistically significant, \update{DAC} took less mean communication time than \update{DSP} ($M = 9.15s$) and \update{DAP} ($M = 10.15s$).

\textbf{Competitive Tasks:} Although the tasks were competitive, we observed participants from three groups communicating for a brief amount of time while performing tasks through \update{SSP} technique. One-way ANOVA showed significant differences among the techniques ($F_{4, 195} = 5.47$, $p < 0.001$, $\eta^2 = 0.10$). Consequently, \update{SSP} technique required communication ($M = 0.40s$), while no communication was observed for all the other techniques (all $M = 0.00s, p = 0.003$) for competitive tasks.

\subsubsection{Annotation}
\textbf{Collaborative Tasks:} Results from the one-way ANOVA showed no significant difference among the techniques in times spent for annotating virtual objects while performing collaborative tasks ($F_{4, 195} = 0.34$, $p = 0.850$, $\eta^2 = 0.01$). Participants spent a brief amount of time annotating virtual objects across all the techniques in a similar pattern: \update{SSP} ($M = 5.43s$), \update{DSP} ($M = 4.15s$), \update{DAP} ($M = 5.03s$), \update{DSC} ($M = 5.45s$), and \update{DAC} ($M = 5.35s$).

\textbf{Competitive Tasks:} Similar to collaborative tasks, no significant differences were found among the techniques for annotating virtual objects during competitive tasks ($F_{4, 195} = 1.29$, $p = 0.277$, $\eta^2 = 0.03$). However, results indicated participants spent less time on average with \update{DAC} ($M = 0.35s$) than all the other techniques: \update{SSP} ($M = 1.83s$), \update{DSP} ($M = 1.10s$), \update{DAP} ($M = 0.70s$), and \update{DSC} ($M = 0.95s$).

\subsubsection{Manipulation}
\textbf{Collaborative Tasks}: Results from the one-way ANOVA showed that there is a significant difference among the techniques in times spent for manipulating virtual objects while performing collaborative tasks ($F_{4, 195} = 3.22$, $p = 0.014$, $\eta^2 = 0.06$). Post-hoc pairwise tests showed no significant differences between the techniques except for \update{SSP} and \update{DSP} ($p = 0.05$). However, \update{SSP} ($M = 9.57s$) and \update{DAC} ($M = 10.25s$) took lesser manipulation time on average than \update{DSP}  ($M = 15.25s$), \update{DAP} ($M = 13.63s$), and \update{DSC} ($M = 14.33s$).

\textbf{Competitive Tasks:} We found a significant difference among the techniques in times spent for manipulating virtual objects while performing competitive tasks ($F_{4, 195} = 14.37$, $p < 0.001$, $\eta^2 = 0.23$). Post-hoc pairwise tests showed that \update{SSP} required significantly higher manipulation time than all other techniques ($p < 0.001$). Noticeably, \update{DAC} ($M = 14.53s$) required less time for manipulation on average than all the other techniques: \update{SSP} ($M = 33.33s$), \update{DSP} ($M = 23.80s$), \update{DAP} ($M = 17.15s$), and \update{DSC} ($M = 20.50s$).

\section{DISCUSSION}
\update{In this paper, we investigated the viewing attributes of MAR user interfaces tailored for remote collaborative tasks, recognizing their critical importance in such contexts. Moreover, we emphasize incorporating features essential for MAR collaboration, including spatial awareness enhancements facilitated by a secondary small screen for real-time view sharing with remote collaborators. In addition to implementing common AR features like viewing the real world while in AR, virtual object manipulation, and interaction, we enabled users to move around the marker to explore models from various perspectives, aligning with the principles of interacting with virtual and real-world objects in AR. Furthermore, we contextualize the potential applications of our system, envisioning its utilization in scenarios such as remote learning/monitoring, where educators/engineers can seamlessly share content with others while monitoring their progress and fostering collaborative discussions.}

Our study examined the influence of different design attributes on remote collaborative and competitive tasks. We designed five techniques with MAR using smartphones and conducted a user study to highlight users' performances, preferences, and behavioral patterns with the techniques. Below, we discuss the key findings from our analysis.

\subsection{Task Performance \update{and Preference}} Our study results indicate that the participants performed better with the techniques -- \update{DAP} \update{(Different Angle - Asynchronous Model - \update{Personal} view)} and \update{DAC} \update{(Different Angle - Asynchronous Model - \update{Combined views})} that support asynchronous model manipulations across collaborative and competitive tasks. While these manipulations were asynchronous, synchronous annotations enabled participants to understand their counterparts' actions better, supported by frequent visual communication between them. Consequently, participants felt lesser mental, physical, and temporal demands with \update{DAP} and \update{DAC}, thus performing better with less effort and frustration. Both participants could simultaneously work on the same task to validate their answers with less negotiation for control over the virtual object. 
Furthermore, with asynchronous model manipulations, participants could independently zoom and rotate the object, enhancing their interaction with the object and comprehension of the other participant's actions. 
Interestingly, the results from the usability and overall preference analysis indicate that \update{DAC} was the most favored technique among the participants which has a small screen for viewing the other participant's activity. This suggests that autonomy in manipulating the shared virtual object and observing the partner’s progress independently enhances task efficiency, usability, and user satisfaction while lowering the perceived workload.

Furthermore, \update{SSP} \update{(Same Angle - Synchronous Model - \update{Personal} View)}, which was designed with synchronous model manipulation and viewing angle between the participants, required the most workload, resulting in poor performance and preference. This is due to the participants being required to negotiate for control of the virtual model frequently and follow how the other person is manipulating the object. Since the point of view of each participant is different from the other and they cannot explicitly see how the other person is holding their mobile phone and AR marker, collaborating on such a situation is much cumbersome. In remote contexts, unlike co-located scenarios, it is important to have seamless object manipulation and a clear strategy to control the ownership of control between each other. We observed that, during collaborative tasks, groups often communicated to assign manipulation responsibilities to one person, while both observed the model to complete the task efficiently.

\subsection{The Impact of \update{a} Secondary Small Screen} The small screen made a noticeable impact on the user preferences for all techniques we designed. Even though the techniques with asynchronous model manipulations performed best during the study (\update{DAP} and \update{DAC}), the overall user preference is for the \update{DAC} technique, where a user can see the other user's activities. During the remote scenarios, the small view emerged as an interestingly popular feature. Similar results were reported in \cite{10.1145/2642918.2647372} where they used small views for remote collaborations between users using tablets and desktop computers. Unlike in the \update{SSP}, \update{DSP} \update{(Different Angle - Synchronous Model - \update{Personal} view)} and \update{DAP} techniques, where the participants can only keep track of their own activities and focus on their own model manipulations, having a dedicated separate screen to see the progress and interactions of the other participants made it a more interactive experience in \update{DSC} \update{(Different Angle - Synchronous Model - \update{Combined views})} and \update{DAC} techniques. During the study, we observed that in some scenarios where one participant could not locate an object, for example, a car in the city model, the other participant located it on his large screen and asked the participant to follow his instructions and observe the small screen to locate the object. Thus, the small screen allowed participants to monitor their partner's actions without direct manipulation interference, bridging an important gap between collaborative practices.

\subsection{Focus on Individual Activities}
 The user preference for techniques that support independent manipulation and the observation of others' progress highlights the necessity of facilitating individual activities within a collaborative framework. Our findings emphasized the need for providing collaborative experiences that happen in real-time and respecting each individual's ability to work independently. To navigate this balance effectively, MAR solutions need to be well designed, ensuring that they cater to the demands of collaborative and competitive interactions. To enhance more collaborative strategies between remotely located participants, it is crucial to design interfaces that offer participants the flexibility to explore and manipulate shared virtual objects independently while also providing mechanisms such as a separate screen - for each participant to remain connected with the overall group's collaborative strategies. 

\subsection{Behavioral Patterns}
We examined the participants’ behavior and strategies for performing collaborative and competitive tasks by analyzing their behavior over communication, annotation, and manipulation. For the collaborative tasks, we observed that the participants spent less time communicating with their partners when they had separate viewing angles. This helped them focus on the virtual object from an angle suitable for themselves. Furthermore, among the techniques with different viewing angles, they completed the assigned task by communicating the least where they could manipulate the virtual object independently and, at the same time, see their partner's activities on the small screen. This happened because verbal communication was replaced by visual communication through the small screen. A similar trend was reported in \cite{10.1145/3555607}.  On the contrary, while performing competitive tasks, participants completely focused on their tasks resulting in no need for communication with their partners at all when they had different viewing angles and could manipulate the virtual object independently. However, they had to communicate when they were sharing the same viewing angle and had to negotiate to take control over the virtual object manipulation.

\update{Participants spent less time annotating virtual objects consistently across techniques, with the shortest annotation time seen when they had the option for independent manipulation and a small screen to monitor partners. This pattern persisted in both collaborative and competitive tasks. Notably, the \update{SSP} technique (same viewing angle, no independent manipulation, and no small screen) and the \update{DAC} technique (different viewing angle, independent manipulation, and small screen) both minimized object manipulation time during collaborative tasks. In \update{SSP}, participants often had one person annotate and manipulate while the other monitored. They had to adopt this strategy because they could not have comfortable viewing angles and needed to negotiate for control whenever they wanted to manipulate the object. This made \update{SSP} the most time-consuming in competitive tasks. On the other hand, \update{DAC}’s support for independent manipulation and the motivational small screen reduced time spent across all behaviors—communication, annotation, and manipulation—making it preferred for efficiency in both task types.}

\subsection{Design Recommendations}
We suggest the following recommendations aim to refine the design of future MAR applications for remote collaboration and competition, ensuring users can interact more seamlessly and effectively. 

\textbf{Viewing Angle:} Our data analysis demonstrates that the participants had to put significant effort into determining stable viewing angles when sharing the same perspective during collaborative tasks (e.g., \update{SSP}) with frequent negotiations for object ownership, this experience was often cumbersome. In remote settings, with absent physical co-location presence, ensuring ease of mutual understanding and cooperation is critical. Similarly, in competitive tasks, we found that task completion times were significantly shorter when participants could view the virtual model from their own viewing angle, compared to when they needed to control the rotation angle each time they interacted with the model. Therefore, we identified it was important to provide mechanisms for participants to control and use their own viewing angle to give a smooth remote experience.

    \textbf{Secondary Small Screen:} Our results showed that it is effective to have a small view in MAR to create activity awareness for performing remote tasks. Results also revealed a strong preference among participants for having this secondary small screen to see what their partner is doing when working on the same virtual model. This suggests a valuable enhancement for future MAR collaborative solutions, highlighting the utility of augmenting a user's personal view with small views of the other users to facilitate a more informed collaborative experience.

   \textbf{User-Centered Asynchronous Manipulation:} We found that task completion times were shorter when participants could manipulate their own virtual model, compared to when they have synchronous manipulation. We thus suggest allowing users to have their own manipulation of the virtual models from their own perspective. This facilitates quicker task completion times and enhances user experience.

\section{LIMITATIONS AND FUTURE WORK}

Our study focused on exploring how different design attributes impact MAR-based remote activities. Although this study provides a set of insights and design guidelines that serve as a foundation for the further development of MAR-based remote activities, it has certain limitations as well. This section discusses those limitations, highlighting the potential scope for future research.

We observed that manipulating 3D models along the x, y, and z axes with 2D touch input is challenging. Consequently, we narrowed our focus to manipulating the rotation of virtual objects centering on AR markers and physically moving the phone around the model. Thus, it warrants further investigation into different 3D object manipulation techniques (i.e., rotating) in remote contexts. 
In the AR application, we used a paper-based marker-based solution, where users can perform object manipulations (e.g., rotation) by moving the marker. Exploring how marker-less AR solutions (e.g., AR Core \cite{ARCore}) would impact the users' experience of remote cooperation would be a future venue of research. Moreover, a significant future investigation could focus on how various form factors, such as the display size of the mobile device and the size of the small screen, impact users' experiences in remote cooperative settings.

\update{We conducted the user study with 20 participants by recruiting them from various ages (between 21 and 34 years) and gender groups (9 female and 11 male). Our sample size aligns with the sample size used in similar HCI studies \cite{10.1145/3313831.3376541, Wells2022:Co-Located_Collaboration}. Similar to Wells et al. \cite{Wells2022:Co-Located_Collaboration}, we used two 3D models to investigate the effect of viewing attributes during remote tasks. We also followed their methodology to place two persons per group and formed 10 groups for the collaborative tasks. Furthermore, prior works \cite{10.1145/3512928, 6104042, 10.1145/3025453.3025537} demonstrated that pairs are sufficient for evaluating general collaboration styles and strategies. While we contend that our sample participants yielded valuable insights, more diversity could be achieved by including a large number of participants from various ages and backgrounds that could reveal additional considerations for future cooperative MAR application designs.}

 \update{The paper aims to explore how participants respond to different viewing attributes instead of focusing on complicated tasks.} Considering the goal of our study, the selected tasks needed to be possible both as an independent task and a collaborative task. \update{Thus, we followed prior works in MAR \cite{10.1145/3313831.3376541, Wells2022:Co-Located_Collaboration} while designing the study scenarios, which asked participants to perform abstract tasks such as counting colored objects. Therefore, we caution that these results may not be generalized as standardized methods for addressing various complex problems in MAR collaboration.} 
 As annotations were related to the virtual object (not a real object), the exact position of this experiment on Milgram’s Reality-Virtuality Continuum \cite{milgram1994taxonomy,10.1177/154193129904302202} is an interesting question - which requires further investigation, particularly in relation to collaborative work. 
As our research used two simple models, it leaves room for future exploration into more complex models (e.g., 3D models of engineering machinery) and varied remote scenarios (e.g., remote cooperative assistance on assembling engineering machinery). 
Furthermore, it would also be interesting to explore remote cooperative activities requiring more than two participants.

\update{The DAP technique involves solving together without sharing, while the DAC method involves solving together while sharing. This makes the study results seem somewhat predictable. However, note that the DAC method requires reserving valuable screen space to the small screen, thereby preventing users from interacting with the entire screen. In contrast, DAP allows users to utilize the entire screen to complete tasks. Consequently, there exists a trade-off between having partial/full-screen space and solving tasks together with/without sharing. However, we recognize the necessity of designing and evaluating a variety of tasks in the MAR context to gain a more comprehensive understanding of how these techniques perform across different scenarios - leading to more insightful conclusions.}

\update{We opted to follow a procedure similar to that used in studies \cite{10.1145/3313831.3376541, Wells2022:Co-Located_Collaboration}, wherein participants shared their answers with an experimenter during the study to receive confirmation. Given the similarity of our study to the papers above, particularly in aspects such as MAR for group collaboration, we decided to follow their design. However, we acknowledge that answers and timing could be recorded automatically for future studies.}
Participants were required to hold their mobile phones pointing at the marker image for a prolonged time during the study - which may trigger arm fatigue. This aspect was not examined in the study due to the simplistic and abstract nature of the tasks involved. Nonetheless, future research could delve into the implications of hand fatigue resulting from prolonged use of mobile handheld AR. Investigating this issue could provide essential insights into ergonomic design principles for MAR applications.

\section{CONCLUSION}
While prior works focused on co-located collaboration using MAR, there is a lack of research exploring various design factors that could impact the overall user experience in remote collaborative and competitive tasks. In this paper, we examined a set of factors, such as content viewing angles to enhance spatial awareness in remote tasks, accommodating different collaboration styles by enabling users to manipulate virtual objects synchronously or asynchronously and incorporating a secondary screen to display each other's activities for creating spatial awareness of their remote activities. we designed a set of techniques incorporating these factors and evaluated them through a user study involving 10 groups of 2 people engaged in remote collaboration and competitive tasks. The results indicated that asynchronous manipulations and having separate views to observe other users' activities led to high performance and better experience in both remote collaborative and competitive activities. Overall, our findings not only enhance user experiences but also open new avenues for remote collaborative and comparative tasks in MAR.


\acknowledgments{%
	This research was funded by an NSERC CREATE grant \# F20-05186.
}

\bibliographystyle{abbrv-doi-hyperref}
\bibliography{main}

\end{document}